\documentclass[aps,pra,preprint,groupedaddress]{revtex4}
\usepackage{graphics}
\begin{document}
\draft

\title{Effects of spatial noncommutativity on energy spectrum of a trapped Bose-Einstein condensate}

\author{You-Hua Luo\footnote{To whom the correspondence should be addressed.\\
 E-mail: yhluo@ecust.edu.cn}, Zi-Ming Ge}

\affiliation{Institute for Theoretical Physics and Department of
Physics, East China University of Science and Technology, Shanghai
200237, China}


\begin{abstract}

In noncommutative space, we examine the problem of a
noninteracting and harmonically trapped Bose-Einstein condensate,
and derive a simple analytic expression for the effect of spatial
noncommutativity on energy spectrum of the condensate. It
indicates that the ground-state energy incorporating the spatial
noncommutativity is reduced to a lower level, which depends upon
the noncommutativity parameter $\theta$. The appeared gap between
the noncommutative space and commutative one for the ground-state
level of the condensate should be a signal of spatial
noncommutativity.

\end{abstract}
\maketitle


The noncommutative geometry plays an important role in string
theory and $M$ theory. For a review on the string theory, see
Ref.~\cite{sei}. Recently there has been a growing interest on the
issue of noncommutative geometry
\cite{dou,cha,gam,nai,sma,pmho,bel,bol,mut,chr,gho,kij,ber,zha1,zha2,zha3,yin,jin,zha4,zha5,bem}.
It is motivated by the discovery in string theory that the low
energy effective theory of D-brane with a non-zero Neveu-Schwarz
(NS)-NS $B$ field background lives on a noncommutative space
\cite{con,douh}. Although the effects of spatial noncommutativity
should presumably become significant at very high energy scales,
e.g., close to the string scale, it is fascinating to speculate
whether there might be some low energy effects of the fundamental
quantum field theory, and it is expected that some low energy
relices of such effects may be verified by current experiments
\cite{zha1}.

In this letter, we examine the effects of spatial noncommutativity
on energy spectrum of a noninteracting and harmonically trapped
Bose-Einstein condensate at the level of quantum mechanics in
noncommutative space (NCQM), in which noncommutativity of both
space-space (space-time noncommutativity is not considered) and
momentum-momentum is considered.

In the case of simultaneously space-space and momentum-momentum
noncommutativity, the consistent deformed Heisenberg-Weyl algebra
\cite{zha1,zha2,zha3,yin} are
\begin{eqnarray}\label{eq:xpij}
  \nonumber  [\hat{x}_{i},\hat{x}_{j}] &=&  i\xi^{2}\theta\varepsilon_{ij}\;,\\
            \ [\hat{p}_{i},\hat{p}_{j}] &=& i\xi^{2}\eta\varepsilon_{ij}\;,\ (i,j=1,2)\\
  \nonumber \ [\hat{x}_{i},\hat{p}_{j}] &=& i\hbar\delta_{ij}\;,
\end{eqnarray}
where $\theta$ and $\eta$ are the noncommutative parameters,
independent of position and momentum; their dimensions are,
$L^{2}$ and $M^{2}L^{2}T^{-2}$, respectively. $\varepsilon_{ij}$
is an antisymmetric unit tensor,
$\varepsilon_{12}=-\varepsilon_{21}=1,
\varepsilon_{11}=\varepsilon_{22}=0$. And $\xi$ is the scaling
factor, it depends upon the noncommutative parameters $\theta$ and
$\eta$.

One possible way of implementing algebra Eqs.~(\ref{eq:xpij}) is
to construct the noncommutative variables
$\{\hat{x}_{i},\hat{p}_{i}\}$ from the corresponding commutative
variables $\{\hat{x}'_{i},\hat{p}'_{i}\}$ through the following
linear transformations:
\begin{equation}\label{eq:transformation}
     \hat{x}_{i} =\xi\left(\hat{x}'_{i}-\frac{\theta\varepsilon_{ij}}{2\hbar}\hat{p}'_{j}\right)\;,
                  \ \ \ \ \ \
     \hat{p}_{i}
     =\xi\left(\hat{p}'_{i}+\frac{\eta\varepsilon_{ij}}{2\hbar}\hat{x}'_{j}\right)\;,
\end{equation}
where $\hat{x}_{i}'$ and $\hat{p}_{i}'$ satisfy the undeformed
Heisenberg-Weyl algebra: $[\hat{x}_{i}',\hat{x}_{j}']=0$,
$[\hat{p}_{i}',\hat{p}_{j}']=0$,
$[\hat{x}_{i}',\hat{p}_{j}']=i\hbar\delta_{ij}$. One can easily
verify that the two first commutation relations of
Eqs.~(\ref{eq:xpij}) can be obtained from
Eqs.~(\ref{eq:transformation}). And the last one is changed to
\begin{equation}\label{eq: xp}
    [\hat{x}_{i},\hat{p}_{j}]=i\hbar\xi^{2}\left(1+\frac{\theta\eta}{4\hbar^{2}}\right)\delta_{ij}\;.
\end{equation}
The Heisenberg commutation relation $[\hat{x}_{i},\hat{p}_{j}]=
i\hbar\delta_{ij}$ should be maintained by
Eqs.~(\ref{eq:transformation}), thus, we identify
$\xi\equiv(1+\frac{\theta\eta}{4\hbar^{2}})^{-1/2}$.

\emph{Bose-Einstein statistics for the case of both
position-position noncommutativity and momentum-momentum
noncommutativity is guaranteed by the deformed Heisenberg-Weyl
algebra itself, independent of dynamics. The deformed bosonic
algebra constitutes a complete and closed algebra}. This theorem
has, very recently, been proved by Zhang \emph{et al.}
\cite{zha5}. The general representations of deformed annihilation
and creation operators $\hat{a}_{i}$ and $\hat{a}_{i}^{\dagger}$
can be constructed from the deformed Heisenberg-Weyl algebra
itself, which reads \cite{zha5}
\begin{equation}\label{eq:aadagger}
 \begin{array}{c}
 \hat{a}_{i}=\sqrt{\frac{1}{2\hbar}\sqrt{\frac{\eta}{\theta}}}\left(\hat{x}_{i}+
     i\sqrt{\frac{\theta}{\eta}}\hat{p}_{i}\right)\;,\\
 \hat{a}_{i}^{\dagger}=\sqrt{\frac{1}{2\hbar}\sqrt{\frac{\eta}{\theta}}}\left(\hat{x}_{i}-
     i\sqrt{\frac{\theta}{\eta}}\hat{p}_{i}\right)\;, \\
 \end{array}
\end{equation}
it follows that the deformed bosonic algebra of $\hat{a}_{i}$ and
$\hat{a}_{j}^{\dagger}$ can be written as
\begin{equation}\label{eq:dba}
     [\hat{a}_{i},\hat{a}_{j}^{\dagger}]=\delta_{ij}+\frac{i\xi^{2}\sqrt{\theta\eta}}{\hbar}\varepsilon_{ij}\;,
                  \
    [\hat{a}_{i},\hat{a}_{j}]=[\hat{a}_{i}^{\dagger},\hat{a}_{j}^{\dagger}]=0\;.
\end{equation}
Eqs.~(\ref{eq:dba}) constitute a complete and closed deformed
bosonic algebra, where the four equations
$[\hat{a}_{1},\hat{a}_{1}^{\dagger}]=[\hat{a}_{2},\hat{a}_{2}^{\dagger}]=1,
[\hat{a}_{1},\hat{a}_{2}]=[\hat{a}_{1}^{\dagger},\hat{a}_{2}^{\dagger}]=0$
are the same as the undeformed bosonic algebra in commutative
space, while the equation
$[\hat{a}_{1},\hat{a}_{2}^{\dagger}]=\frac{i}{\hbar}\xi^{2}\sqrt{\theta\eta}$
is the new one. It codes effects of spatial noncommutativity.
Inserting Eqs.~(\ref{eq:transformation}) into
Eqs.~(\ref{eq:aadagger}), we can obtain the relations between the
deformed annihilation and creation operators
$\{\hat{a}_{i},\hat{a}_{i}^{\dagger}\}$ and the undeformed
annihilation and creation operators
$\{\hat{a}_{i}',\hat{a}_{i}'^{\dagger}\}$:
\begin{equation}\label{eq:aa}
 \begin{array}{c}
 \hat{a}_{i}=\xi(\hat{a}_{i}'+\frac{i}{2\hbar}\sqrt{\theta\eta}\varepsilon_{ij}\hat{a}_{j}')\;,\\
 \ \hat{a}_{i}^{\dagger} = \xi(\hat{a}_{i}'^{\dagger}-\frac{i}{2\hbar}\sqrt{\theta\eta}\varepsilon_{ij}\hat{a}'^{\dagger}_{j})\;. \\
 \end{array}
\end{equation}

For the condensate consisting of $N$ noninteracting bosons, the
Hamiltonian can be written as
\begin{equation}\label{eq:Hamiltonian}
   \hat{H}=\sum_{i=1}^{N}\left(-\frac{\hbar^{2}}{2m}\bigtriangledown_{i}^{2}+V_{\mathrm{trap}}(\vec{r}_{i})\right)\;,
\end{equation}
in which includes the kinetic energy of atoms and the potential
energy due to the trap. An important feature characterizing the
available magnetic traps for alkali atoms is that the confining
potential can be safely approximated with the quadratic form
\cite{dal}
\begin{equation}\label{eq:trap}
  V_{\mathrm{trap}}(\vec{r})=\frac{1}{2}m(\omega_{x}^{2}x^{2}+\omega_{y}^{2}y^{2}+\omega_{z}^{2}z^{2})\;.
\end{equation}
In this case of neglecting the atom-atom interaction, almost all
predictions are analytical and relatively simple. The many-body
Hamiltonian is the sum of single-particle Hamiltonians whose
eigenvalues have the form:
$\epsilon_{n_{x}n_{y}n_{z}}=(n_{x}+\frac{1}{2})\hbar\omega_{x}+(n_{y}+\frac{1}{2})\hbar\omega_{y}+
(n_{z}+\frac{1}{2})\hbar\omega_{z}\;,$ where
$\{n_{x},n_{y},n_{z}\}$ are non-negative integers. The
ground-state $\phi(\vec{r}_{1},\cdot\cdot\cdot,\vec{r}_{N})$ of
$N$ noninteracting bosons confined by the potential
(\ref{eq:trap}) is obtained by putting all the particles in the
lowest single-particle (SP) state $(n_{x}=n_{y}=n_{z}=0)$, namely
$\phi(\vec{r}_{1},\cdot\cdot\cdot,\vec{r}_{N})=\prod_{i}\varphi_{0}(\vec{r}_{i})$.
And the corresponding ground-state energy spectrum of the
condensate
$E=N\epsilon_{000}=\frac{1}{2}N\hbar(\omega_{x}+\omega_{y}+\omega_{z})\;$.

In noncommutative space, for simplicity, we restrict ourselves to
the case of the Bose-Einstein condensate (BEC) in a
two-dimensional isotropic harmonic trap, and choose the axis of
rotation to be the $z$ axis, and we assume that the ultra-cold
bosonic atoms are in the their ground state along the $z$
direction, thus solving an effectively two-dimensional problem.
For a noninteracting and harmonically trapped BEC, our starting
point is the SP Hamiltonian
\begin{equation}\label{eq:hamt}
    \hat{H}=\frac{1}{2m}(\hat{p}_{x}^{2}+\hat{p}_{y}^{2})+
    \frac{1}{2}m\omega^{2}(\hat{x}^{2}+\hat{y}^{2})\;.
\end{equation}
Here $m$ is the mass of the atoms, and the trapping potential is
assumed to be an isotropic harmonic oscillator of frequency
$\omega$. In the above case, the condition of guaranteeing
Bose-Einstein statistics reads
\begin{equation}\label{eq:condition}
    \frac{1}{m\omega}=\sqrt{\frac{\theta}{\eta}}\;.
\end{equation}
Equation (\ref{eq:condition}) yields a direct proportionality
between the two noncommutative parameters.

For clarity, we identified $\hat{x}_{1}\equiv\hat{x},\
\hat{x}_{2}\equiv \hat{y}, \ \hat{p}_{1}\equiv\hat{p}_{x},\
\textrm{and}\ \hat{p}_{2}\equiv\hat{p}_{y}\;$. The SP Hamiltonian
can be actually decomposed two independent harmonic oscillators,
namely,
\begin{equation}\label{eq:harmos}
   \hat{H}=\hbar\omega\left(\hat{a}_{1}^{\dagger}\hat{a}_{1}+\frac{1}{2}\right)+
   \hbar\omega\left(\hat{a}_{2}^{\dagger}\hat{a}_{2}+\frac{1}{2}\right)\;.
\end{equation}
In terms of Eqs.~(\ref{eq:aa}), we can obtain the Hamiltonian
including the noncommutative parameter $\theta$, which is given by
\begin{widetext}
\begin{equation}\label{eq:hamtseta}
    \hat{H}=\hbar\omega\left\{\xi^{2}\left[1+\left(\frac{m\omega\theta}{2\hbar}\right)^{2}\right]\left(\hat{a}'^{\dagger}_{1}\hat{a}'_{1}+
    \hat{a}'^{\dagger}_{2}\hat{a}'_{2}\right)+\frac{im\omega\theta}{\hbar}\xi^{2}\left(\hat{a}'^{\dagger}_{1}\hat{a}'_{2}-
    \hat{a}'^{\dagger}_{2}\hat{a}'_{1}\right)+1\right\}.
\end{equation}
\end{widetext}
In order to diagonalize the above Hamiltonian, we employ
like-Bogoliubov's transformation:
\begin{equation}\label{eq:btf}
     \hat{b}_{i}^{\dagger}=u\hat{a}'^{\dagger}_{i}-v\hat{a}'^{\dagger}_{j}\;,
                  \ \ \ \ \ \
      \hat{b}_{i}=u\hat{a}'_{i}+v\hat{a}'_{j}\;, \ (i,j=1,2)
\end{equation}
where $u^{2}-v^{2}=1$, and $\hat{b}_{i}, \hat{b}_{j}^{\dagger}$
meet the usual bosonic commutation relations
$[\hat{b}_{i},\hat{b}_{j}^{\dagger}]=\delta_{ij}, \
[\hat{b}_{i},\hat{b}_{j}]=[\hat{b}_{i}^{\dagger},\hat{b}_{j}^{\dagger}]=0$.
Thus, Eq.~(\ref{eq:hamtseta}) can be rewritten as
\begin{widetext}
\begin{eqnarray}
 \nonumber  \hat{H} &=& \hbar\omega+\hbar\omega\xi^{2}\left[1+\left(\frac{m\omega\theta}{2\hbar}\right)^{2}-
   2uv\frac{im\omega\theta}{\hbar}\right]\hat{b}_{1}^{\dagger}\hat{b}_{1}+\hbar\omega\xi^{2}\left[1+
   \left(\frac{m\omega\theta}{2\hbar}\right)^{2}+2uv\frac{im\omega\theta}{\hbar}\right]\hat{b}_{2}^{\dagger}\hat{b}_{2} \\
  & & +\hbar\omega\left(u^{2}+v^{2}\right)\frac{im\omega\theta}{\hbar}\xi^{2}\left(\hat{b}_{1}^{\dagger}\hat{b}_{2}-
   \hat{b}_{2}^{\dagger}\hat{b}_{1}\right)\;.
\label{eq:dhamt}
\end{eqnarray}
\end{widetext}
Let $u^{2}+v^{2}=0$, and combine with $u^{2}-v^{2}=1$, we can
obtain $u=\pm\sqrt{2}/2, v=\pm i\sqrt{2}/2$ . The resulting
Hamiltonian
\begin{widetext}
\begin{equation}\label{diah}
    \hat{H}=\hbar\omega+\hbar\omega\xi^{2}\left[1+\left(\frac{m\omega\theta}{2\hbar}\right)^{2}+
    \frac{m\omega\theta}{\hbar}\right]\hat{b}_{1}^{\dagger}\hat{b}_{1}+\hbar\omega\xi^{2}\left[1+
    \left(\frac{m\omega\theta}{2\hbar}\right)^{2}-
    \frac{m\omega\theta}{\hbar}\right]\hat{b}_{2}^{\dagger}\hat{b}_{2}\;.
\end{equation}
\end{widetext}
It shows that the corresponding energy spectrum of SP is
$e_{0}=\hbar\omega$,
$e_{1}=e_{0}+m\omega^{2}\theta+O(\theta^{3})$, and
$e_{2}=e_{0}-m\omega^{2}\theta+O(\theta^{3})$, respectively. In
commutative space, the ground-state level of BEC is
$N\hbar\omega$, while in noncommutative space, its ground-state
level is changed to $N\hbar\omega-Nm\omega^{2}\theta$. It
indicates that the ground-state energy incorporating the spatial
noncommutativity is reduced to a lower level. Namely, There is an
energy gap between the noncommutative space and commutative one
for the ground-state level of the condensate, its value is
$\Delta=Nm\omega^{2}\theta$. It should be a signal of spatial
noncommutativity.

For the order of magnitude of the gap above-mentioned, we can make
a rough estimate. On the one hand, the size of ultra-cold atomic
cloud is instead independent of $N$ and is fixed by the harmonic
oscillator length \cite{dal}:
$a_{\mathrm{ho}}=(\frac{\hbar}{m\omega_{\mathrm{ho}}})^{1/2}$. In
the available experiments, it is typically of the order of
$a_{\mathrm{ho}}\approx 1\ \mathrm{\mu m}$. One the other hand,
assuming $1\times10^{5}$ noninteracting ultra-cold $^{87}$Rb atoms
in a harmonic trap, namely, in the expression of the gap,
$N=1\times10^{5}$, $m\doteq1.5\times10^{-25}$ kg, and
$\omega\doteq 7.0\times10^{2}\ \mathrm{s^{-1}}$. For the value of
parameter $\theta$, the existing experiments \cite{car} give
$\theta/(\hbar c)^{2}\leq(10 \ \mathrm{TeV})^{-2}$, $\theta\simeq
4\times 10^{-40}\ \mathrm{m}^{2}$. Therefore, the estimated gap
$\Delta\doteq1.8\times10^{-35}\ \mathrm{eV}$. Obviously, the order
of magnitude of the gap is extremely small. It may be difficult to
detect effects of spatial noncommutativity by means of present
experimental technology. However, if we take the atom-atom
interaction into account in BEC, and squeeze the size of
ultra-cold atomic cloud, the order of magnitude of the gap can be
increased, and make it possible to detect effects of spatial
noncommutativity.

In summary, in noncommutative space, we examine the problem of a
noninteracting and harmonically trapped Bose-Einstein condensate,
and derive a simple analytic expression for the effect of spatial
noncommutativity on energy spectrum of the condensate. We find
that there is an energy gap between the noncommutative space and
commutative one for the ground-state level of the condensate.
Also, we show that the order of magnitude of the gap is extremely
small $(\sim 10^{-35}\ \mathrm{eV})$. At present, it may be
difficult to detect effects of spatial noncommutativity.


\begin{acknowledgments}
Y.-H. Luo greatly appreciates Prof. J.-Z. Zhang for his help and
discussions. This work is supported in part by the National
Natural Science Foundation of China under the grant number
10174086.
\end{acknowledgments}

\end{document}